\documentclass[12pt]{article}
\usepackage[english]{babel}

\begin{document}

\title{Degeneracy of Schr$\ddot{o}$dinger equation with potential $1/r$ in $d$-dimensions}

\author{M. A. Jafarizadeh $^{1, 2}$ , H. Goudarzi$^{1,3}$\footnote{E-mail: Goudarzia@phys.msu.ru} \\ 
{\footnotesize \textit{1. Department of Theoretical Physics, Faculty of Physics, Tabriz University, Tabriz 51664, Iran}}\\ \textit{{\footnotesize 2. Institute for Studies in Theoretical Physics and Mathematics, Tehran, Iran}}\\ \textit{{\footnotesize 3. Department of Physics, Oroumia University, Oroumia, Iran}}}

\date{}
\maketitle 

{\footnotesize{\textit{Received 18 March 1997, accepted 24 September 1997}}}\\

\begin{abstract}
{\footnotesize {Using the irreducible representations of the group $SO(d+1)$, we discuss the degeneracy symmetry of hydrogen atom in $d$-dimensions and calculate its energy spectrum as well as the correspouding degeneracy. We show that $SO(d+1)$ is the energy spectrum generating group.}}
\\
\end{abstract}
{\footnotesize{PACS}} : \  03.65.Fd \\ \\
{\footnotesize{KEYWORDS}} : {\small{Schrodinger equation, degeneracy, energy levels}} \\ \\

\begin{center}
\title\textbf{1. Introduction}
\maketitle
\end{center}

There is a wealth of references concerning calculations of energy spectrum and degeneracy of Schrodinger equation with potential equal $1/r$ (i.e. hydrogen atom) in literature [1]. Almost all of them are confined within the limits of our observed world. However, it is a common practice to consider $(d+1)$ dimensional space-time, e.g. in the domain of string theory [2] or the Kaluza-Klein theories [3]. We generalize the matter upto (spatial) $d$-dimensions and evaluate the energy spectrum. Symmetry plays an important role in calculating the eigenstate of a Hamiltonian. Symmetry and degeneracy of energy levels of a system are inter-related [4-7].\\

In Section 2., of the present paper, we show that the group $SO(d+1)$ is the degeneracy group of the $d$-dimensional Schrodinger equation with potential $1/r$. By introducing $d(d+1)/2$ generators as the generators of the $SO(d+1)$ algebra which satisfy the commutation relations of the algebra, we show that the Hamiltonian of the system is invariant under the group $SO(d+1)$, and that the Casimir operator of the $SO(d+1)$ algebra gives its spectrum, and also that the degeneracy number for a given energy is the dimension of the irreducible representation of that group.

In Section 3., we introduce the hyperspherical harmonics which are themselves the irreducible representations of the rotation group in $d$-dimensions, i.e. $SO(d)$. Next in Section 4., the $d$-dimensional Scherodinger equation in hyperspherical coordinates are calculated with the aid of these functions. The derived energy spectrum is also compared with the result obtained in Sectoin 2.\\
  
\begin{center}
\title\textbf{2. Schr$\ddot{\textbf{o}}$dinger equatoin with potential $\textbf{1/r}$ and\\ degeneracy group $\textbf{SO(d+1)}$}
\maketitle
\end{center}

We solve the Schrodinger equation by using the degeneracy symmetry of the group $SO(d+1)$ in $d$-dimensions and show that it corresponds to that of the analytical solution. This means, we must show that $d$-dimensional Schrodinger equation has an $SO(d+1)$ degeneracy symmetry, with a spectrum as calculated by the Casimir operator of that group. Also we obtain its degeneracy number by finding the irreducible representation of the group $SO(d+1)$.

The generators and the Poissonian brackets of the rotation group $SO(d)$ satisfy the following relations
$$L_{ij}=x_ip_j - x_jp_i \;; \ \ \ \ \ \  \ \ i, j = 1, 2,  \cdots , d.$$
$$\left\{L_{ij}, L_{kl}\right\}=\delta_{jl}L_{ik}+ \delta_{ik}L_{jl}+ \delta_{jk}L_{li} +\delta_{il}L_{kj}.$$

Now, one can easily transform these 'classical' relations into quantum mechanics and hence find the commutation relations. We also note that the quantum mechanical Hamiltonian is the same as the classified one:
$$H=\sum_{i=1}^d \frac{p_i^2}{2\mu}- \frac{k}{r}\;.$$
We remind ourselves that $H$ is invariant under rotation, therefore
\begin{equation}
\frac{d}{dt}L_{ij}=0.
\end{equation}
The quantum mechanical Range - Lenz vector is defined as
$$M_i= \frac{1}{2\mu}\sum_{j=1}^d\left(P_jL_{ij}+L_{ij}P_j\right)- \frac{k}{r}\; x_i \;,$$
where $\mu$ is the reduced mass and $k$ is a constant. Also note that $M_i$ are integrals of motion, that is:
\begin{equation}
\frac{d}{dt}M_i=0.
\end{equation}
So we have 
$$\left[H, L_{ij}\right]=0$$
$$\left[H, M_i\right]=0\;.$$
Also, the commutation relations between $M_i$ and $L_{ij}$ are 
\begin{equation}
\left[M_l, M_k\right]=-i\hbar\frac{2H}{\mu}L_{lk}\;,
\end{equation} 
\begin{equation}
\left[M_i, L_{kl}\right]=-i\hbar \left(\delta_{il}M_k - \delta_{ik}M_l\right).
\end{equation}
Taking eq. (3) into consideration, we introduce generator $M_i^{'}$ as
\begin{equation} 
M_i^{'}=\frac{M_i}{\sqrt{-2H/\mu}}.
\end{equation}
It is clrar that
\begin{equation}
\left[M_l^{'}, M_k^{'}\right]=i\hbar L_{lk}.
\end{equation}
Eqs. (4) and (6) are the commutation relations of the group $SO(d+1)$.\\

Now, in order to calculate the energy spectrum of the Schrodinger equation for the potential $1/r$ in $d$-dimensions, we must first write the Casimir operator for the group $SO(d+1)$:
\begin{equation}
C=L_{ij}^2 + {M_i^{'}}^{2}
\end{equation}
Using the following commutation relations
$$\left[P_j , L_{ij}\right]=i\hbar P_i(d-1) ,$$
$$\left[P_i , L_{ij}\right]=- i\hbar P_j (d-1) ,$$
$$\left[P_k , L_{ij}\right]=-i\hbar P_j\delta_{ki} + i\hbar P_i\delta_{kj}.$$
One can easily verify that
$$M_i^2 = \frac{2H}{\mu}\left[L^2+{\left(\frac{d-1}{2}\right)}^2\right]+k^2 .$$
Hence,
$$C=L_{ij}^2 + {M_i^{'}}^{2}={\left(\frac{d-1}{2}\right)}^2 - \frac{\mu k^2}{2H} ,$$
where we have made use of eq. (5) and
$$\sum_{i<j}L_{ij}^2 = L^2 .$$
The eigenvalue of the Casimir operator $C$ for the group $SO(d+1)$ is
\begin{equation}
C=n(n+d-1)\hbar^2
\end{equation}
Therefore,
\begin{equation}
n(n+d-1)= {\left(\frac{d-1}{2}\right)}^2 - \frac{\mu k^2}{2E_n} ,
\end{equation}
where $E_n$ is the eigenvalue of the Hamiltonian.

Rewriting
$$n(n+d-1) = \left[n+\left(\frac{d-1}{2}\right)\right]^2 - \left(\frac{d-1}{2}\right)^2$$
and substituting this into eq. (9), we obtain
\begin{equation}
	E_n = \frac{-\mu k^2}{2\hbar^2 {\left(n+\frac{d-1}{2}\right)}^2}\;.
\end{equation}

With regard to the commutation relations, where it is explicitly shown that $H$ commutes with all generators of the group $SO(d+1)$, it is quite clear that according to the Schur's lemma [4, 5], the Hamiltonian must somehow be related to the Casimir operator of the group. All quantum eigenstates with energy given by eq. (10) belong to the irreducible represetation of that group with eigenvalue of the Casimir operator given in eq.(8). The degeneracy number is the dimension of the representation which according to eq. (23) of Section 3. is equal to
$$g = \frac{(2n+d-1)(n+d-2)!}{n! (d-1)!} \;.$$ 

\begin{center}
\title\textbf{3. Hyperspherical harmonics in $d$-dimensions}
\maketitle
\end{center}

We demonstrate that Gegenbauer hyperspherical harmonics are the irreducible representations of the group $SO(d)$. Then, using the tensorial representations of the degenerate group $SO(d)$, we calculate the dimension of the representation.\\

The $d$-dimensional Laplacian in hyperspherical coordinates is defined as
\begin{equation}
\nabla^2=-\frac{L^2}{r^2}+\frac{1}{r^{d-1}}\frac{\partial}{\partial r}\left(r^{d-1}\frac{\partial}{\partial r}\right) ,
\end{equation} 
where $L^2$ contains angular components of the Laplacian and $r$ is the radial component in hyperspherical coordinates.

In $d$-dimensional hyperspherical coordinates, we have 

$x_1=r cos\vartheta_1 $ 

$x_2=rsin\vartheta_1 cos\vartheta_2$ 

$\vdots $

$x_{d-1}=r sin\vartheta_1 sin\vartheta_2 \cdots cos\vartheta_{d-1}$

$x_d=r sin\vartheta_1  \cdots sin\vartheta_{d-2} sin\vartheta_{d-1}$\\
with the length element as
$$ds^2= g_{\alpha\beta}dq^\alpha dq^\beta$$
with $q_1=r$ and $q_i=\vartheta_i$, $(i=2, 3, \cdots , d-1)$ and where $g_{\alpha\beta}$, the metric of the space, is defined as
$$g_{\alpha\beta}=diag(1, r^2, r^2 sin^2\vartheta_1, \cdots, r^2 sin^2\vartheta_1 \cdots sin^2\vartheta_{d-2})\;.$$
Writing $L^2$ in hyperspherical coordinate axis, we obtain \\
$$L^2= \frac{1}{sin^{d-2}\vartheta_1}\frac{\partial}{\partial\vartheta_1}sin^{d-2}\vartheta_1\frac{\partial}{\partial\vartheta_1}+ \frac{1}{sin^{2}\vartheta_1 sin^{d-3}\vartheta_2}\frac{\partial}{\partial\vartheta_2}sin^{d-3}\vartheta_2\frac{\partial}{\partial\vartheta_2}+ $$
$$+\frac{1}{sin^{2}\vartheta_1 sin^{2}\vartheta_2 sin^{d-4}\vartheta_3}\frac{\partial}{\partial\vartheta_3}sin^{d-4}\vartheta_3\frac{\partial}{\partial\vartheta_3} + \; \cdots + $$
$$+\frac{1}{sin^2\vartheta_1 sin^2\vartheta_2 \cdots sin^2\vartheta_{d-2}}\frac{\partial^2}{\left(\partial\vartheta_{d-1}\right)^2}\;.$$
One can easily see that $L^2$ satisfies the following recursion relation:
\begin{equation}
L_{(k+1)}^2=-\frac{1}{sin^{k-1}\vartheta_{d-k}}\frac{\partial}{\partial\vartheta_{d-k}}sin^{k-1}\vartheta_{d-k}\frac{\partial}{\partial\vartheta_{d-k}}+ \frac{L_{(k)}^2}{sin^2\vartheta_{d-k}}\;.
\end{equation} 

In order to find the eigenfunctions and the eigenvalues of $L^2$, we benefit from the resemblance with the rotational group $SO(3)$, where its eigenfunctions, i.e. its irreducible representations, are $Y_{lm}(\vartheta, \varphi)$. One can write the eigenvalue relation for $L_{(d)}^2$ as
\begin{equation}
L_{(d)}^2Y_{l_{d-1} l_{d-2} \cdots l_2 l_1}(\vartheta_1, \vartheta_2, \cdots, \vartheta_{d-1})=l_{d-1}(l_{d-1}+d-2)Y_{l_{d-1} l_{d-2} \cdots l_2 l_1}(\vartheta_1, \vartheta_2, \cdots, \vartheta_{d-1}).
\end{equation}
Now, we prove that $Y_{l_{d-1} l_{d-2} \cdots l_2 l_1}(\vartheta_1, \vartheta_2, \cdots, \vartheta_{d-1})$ are the eigenfunctions of $L_{(d)}^2$, that is they are the irreducible representations of the group $SO(d)$ which satisfy equation (13) as well as the following
$$\int d\Omega Y^*_{l_{d-1} l_{d-2} \cdots l_2 l_1}(\vartheta_1, \vartheta_2, \cdots, \vartheta_{d-1})Y_{l^{'}_{d-1} l^{'}_{d-2} \cdots l^{'}_2 l^{'}_1}(\vartheta_1, \vartheta_2, \cdots, \vartheta_{d-1})=$$
$$ = \delta_{l_{d-1}l^{'}_{d-1}} \delta_{l_{d-2}l^{'}_{d-2}} \cdots \delta_{l_2l^{'}_2} \delta_{l_1l^{'}_1}\;,$$
where $Y_{l_{d-1} l_{d-2} \cdots l_2 l_1}(\vartheta_1, \vartheta_2, \cdots, \vartheta_{d-1})$ are the hyperspherical harmonics.\\

In order to find an expression in which the eigenvalues of $L_{(k)}^2$ hold, and to obtain the corresponding differential equation, we write
\begin{equation}
L_{(k+1)}^2 Y_{l_k l_{k-1} \cdots l_1}(\vartheta_1, \cdots, \vartheta_{k-1})=l_k(l_k+k-1)Y_{l_k l_{k-1} \cdots l_1}(\vartheta_1, \cdots, \vartheta_{k-1}), 
\end{equation} 
\begin{equation}
L_{(k)}^2 Y_{l_k l_{k-1} \cdots l_1}(\vartheta_1, \cdots, \vartheta_{k-1})=l_{k-1}(l_{k-1}+k-2)Y_{l_k l_{k-1} \cdots l_1}(\vartheta_1, \cdots, \vartheta_{k-1})\;. 
\end{equation}
From eqs. (12) and (14, 15), we derive the following differential equation:
$$l_k(l_k+k-1)C_{l_k, l_{(k-1)}}^{((k-2)/2)}(cos\vartheta_k)=- \frac{1}{sin^{k-1}\vartheta_{k}}\frac{\partial}{\partial\vartheta_{k}}sin^{k-1}\vartheta_{k}\frac{\partial}{\partial\vartheta_{k}} C_{l_k, l_{k-1}}^{((k-2)/2)}(cos\vartheta_k)+ $$
\begin{equation}
+ \frac{l_{k-1}(l_{k-1}+k-2)}{sin^2\vartheta_{k}}C_{l_k, l_{k-1}}^{((k-2)/2)}(cos\vartheta_k)\;,
\end{equation}
where
$$Y_{l_k l_{k-1} \cdots l_1}(\vartheta_1, \cdots, \vartheta_{k-1})=C_{l_k, l_{k-1}}^{((k-2)/2)}(cos\vartheta_k) Y_{l_{k-1} l_{k-2} \cdots l_1}(\vartheta_1, \cdots, \vartheta_{k-2})\;.$$
Eq. (16) is the most general differential equation in which $C_{l_k, l_{k-1}}^{((k-2)/2)}(cos\vartheta_k)$ are satisfied. To solve this equation, we put
$$x_k=cos\vartheta_k $$
Hence the associated Gegenbauer differential equation [8]:
$$\left[\frac{1}{\left(1-x_k^2\right)^{(k-2)/2}}\frac{d}{dx_k}(1-x_k^2)^{k/2}\frac{d}{dx_k}+ l_k(l_k+k-1)- \frac{l_{k-1}(l_{k-1}+k-2)}{1-x_k^2}\right]\times $$
\begin{equation}
\times C_{l_k, l_{k-1}}^{((k-2)/2)}(x_k)=0
\end{equation}

To solve eq. (17), we consider first the case in which the last term is absent, that is 
\begin{equation}
 \left[\frac{1}{\left(1-x_k^2\right)^{(k-2)/2}}\frac{d}{dx_k}(1-x_k^2)^{k/2}\frac{d}{dx_k}+ l_k(l_k+k-1)\right] C_{l_k, l_{k-1}}^{((k-2)/2)}(x_k)=0\;,
\end{equation}
of which we get the following solution:
$$ C_{l_k, l_{k-1}}^{((k-2)/2)}(x_k)= a_{l_k}\frac{1}{\left(1-x_k^2\right)^{(k-2)/2}}\left(\frac{d}{dx_k}\right)^{l_k}\left[\left(1-x_k^2\right)^{l_k+(k-2)/2}\right]\;.$$
The normalization condition determines the coefficient $a_{l_k}$
$$a_{l_k}=\left[\frac{(k-2)!\Gamma(l_k+k/2+l/2)}{\sqrt{\pi}(2l_k+k-2)!(l_k+k/2-l)!}\right]^{1/2}$$
Now, in order to solve eq. (17), we note that having differentiated eq. (18) $m$ times, where $m=l_{k-l}$, we obtain the following equation:
$$\left(1-x_k^2\right)\frac{d^2}{dx_k^2}C_{l_k}^{(m)}(x_k)+ (-kx_k-2mx_k)\frac{d}{dx_k}C_{l_k}^{(m)}(x_k)+$$
\begin{equation}
+\left[1-m(m-1)-km+l_k(l_k+k-1)\right]C_{l_k}^{(m)}(x_k)=0.
\end{equation}
The solution of eq. (19) can be shown to be
\begin{equation}
C_{l_k}^{(m)}(x_k)=u(x_k)C_{l_k, l_{k-1}}^{((k-2)/2)}(x_k).
\end{equation} 
Now, substituting eq. (20) in eq. (19), we obtain
$$\left(1-x_k^2\right) {C^{''}}_{l_k, l_{k-1}}^{((k-2)/2)}(x_k)+\left[2\frac{u^{'}}{u}\left(1-x_k^2\right)-kx_k-2mx_k\right]{C^{'}}_{l_k, l_{k-1}}^{((k-2)/2)}(x_k)+ $$
$$+\left[\left(1-x_k^2\right)\frac{u^{''}}{u}-(kx_k-2mx_k)\frac{u^{'}}{u}l_k(l_k+k-1)-km-m(m-1)\right]\times $$
\begin{equation}
 \times C_{l_k, l_{k-1}}^{((k-2)/2)}(x_k)=0
\end{equation} 
In order that the differential equation (21) preserves its initial form, i.e. eq. (18), the following relation must hold:
\begin{equation}
	\left(1-x_k^2\right)^2\frac{u^{'}}{u}-2mx_k=0
\end{equation}
From eq. (22) we get
$$u(x_k)=\left(1-x_k^2\right)^{-m/2}$$

Note that differentiating once from eq. (22) with respect to $x$ and applying condition (22) on eq. (21), we get eq. (17). This indicates that the proposed solution (20) is the solution of the equation (17):
\begin{equation}
	C_{l_k}^{l_{k-1}}(x_k)=\gamma \left(1-x_k^2\right)^{(l_{k-1})/2}\left(\frac{d}{dx_k}\right)^{l_{k-1}}C_{l_k}(x_k).
\end{equation}
Orthonormality determines the coefficient $\gamma$ of eq. (23):
$$\gamma=(-1)^ma_{l_k}\frac{(l_k+k+m-2)!}{(l_k+k-m-2)!}\;.$$
Having obtained the general solution of the differential equation (17), now we write down the explicit form of the hyperspherical harmonics as
$$Y_{l_{d-1} \cdots l_1}(\vartheta_1, \cdots, \vartheta_{d-1})= $$
$$= C_{l_{d-1}}^{l_{d-2}}(cos\vartheta_{d-1})C_{l_{d-2}}^{l_{d-3}}(cos\vartheta_{d-2})\cdots C_{l_3}^{l_2}(cos\vartheta_3)C_{l_2}^{l_1}(cos\vartheta_2)C_{l_1}(cos\vartheta_1)$$
which satisfy the orthonormality relation.\\

We complete this section by calculating the dimension of the irreducible representation of group $SO(d)$. To do this we remind ourselves that traceless symmetrical tensors $T_{i1 i2 \cdots il}$ are also irreducible representations of that group. So we calculate the number of permutations of the indices of the tensor $T$. The result is
$$g_l(l_k)=\frac{(l_k+d-1)!}{(d-1)!l_k!}$$
Since the tensors are traceless, therefore the degeneracy number is calculated by the following relation:
\begin{equation}
	g(l_k)=g_l(l_k)-\frac{(l_k+d-3)!}{(d-1)(l_k-2)!}=\frac{(2l+d-2)(l_k+d-3)!}{(d-2)!l_k!} \;,\ \ \  k=d-1
\end{equation}\\

\begin{center}
\title\textbf{4. Solution of the radial Schrodinger equation with potential $1/r$ in $d$-dimensions}
\maketitle
\end{center}

consider the following Schrodinger equation
\begin{equation}
\left(-\frac{\hbar^2}{2\mu}\nabla^2+V(r)\right)\psi(r)=E\psi(r)
\end{equation}
with the central potential defined as
$$V(r)=-\frac{k}{r}\:,$$
where $k$ is a constant and $r$ is the radius of a $d$-dimensional sphere:
$$r= \sqrt{\sum_{i=1}^dx_i^2}$$
with the Laplacian defined by eq. (11). Inserting the Laplacian in eq. (25), we get
\begin{equation}
  \left[-\frac{L^2}{r^2}+\frac{1}{r^{d-1}}\frac{\partial}{\partial t}\left(r^{d-1}\frac{\partial}{\partial t}\right)+\frac{2mk}{\hbar^2r}+\frac{2mE}{\hbar^2}\right]\psi(r)=0
\end{equation}
On separating the variables according as
$$\psi(r)=R(r)Y_{l_{d-1} l_{d-2} \cdots l_2 l_1}(\vartheta_1, \vartheta_2, \cdots, \vartheta_{d-1})$$
and making use of the eigenvalue equation of the spherical harmonics i.e. eq. (13), the differential equation (26) transforms into
\begin{equation}
  R^{''}(r)+\frac{d-1}{r}R^{'}(r)+ \left[\frac{2mk}{\hbar^2r}+\frac{2mE}{\hbar^2}-\frac{l_{d-1}(l_{d-1}+d-2)}{r^2}\right]R(r)=0
\end{equation} 

This is the radial differential equation in $d$-dimensions, by means of which one can calculate the energy spectrum. To do this, we consider first the asymptotic behavior of $R(r)$ :
\begin{equation}
  R(r)=r^\alpha e^{i\beta r}Y_n(r)
\end{equation}
where $Y_n(r)$ are the confluent hypergeometric functions. Substituting eq. (28) into eq. (27) one can see that $Y_n(r)$ satisfy the confluent hypergeometric equation
$$rY_n^{''}(r)+ (2\alpha + 2i\beta r + (d-1))Y_n^{'}+ \textbf{[}\alpha(\alpha -1)\frac{1}{r}+ 2i\alpha \beta - r\beta ^2 + $$
\begin{equation}
+ \alpha (d-1)\frac{1}{r} + i\beta + (d-1)+\frac{2mk}{\hbar^2}+\frac{2mEr}{\hbar^2}- l_{d-1}(l_{d-1}+d-2)\frac{1}{r}\textbf{]}Y_n(r)=0
\end{equation}
We know that the general form of confluent hypergeometric differential equation are of the following form
\begin{equation}
  xY^{''}(x)+(c-x)Y^{'}(x)-aY(x)=0
\end{equation}
In otder that eq. (29) reduces to the standard form (30), the parameters $\alpha$ and $\beta$ must satisfy 
\begin{equation}
	\alpha = l_{d-1}\;, \ \ \ \ \ \ \ \ \beta = \frac{2mE}{\hbar^2}
\end{equation}
With a change in variable as 
$$2i\beta r = -x$$
the eq. (29) becomes
\begin{equation}
  xY^{''}_n(x)+ \left[(2l_{d-1}+d-1)-x\right]Y^{'}_n(x)-\left[l_{d-1}+\frac{d-1}{2}-\frac{imk}{\beta \hbar^2}\right]Y_n(x)=0.
\end{equation}
Now, in order to have a polynomial solution to eq.(32), we must have 
\begin{equation}
	l_{d-1}+ 	\frac{d-1}{2}- \frac{imk}{\beta \hbar^2}=-j ,
\end{equation}
with $j$ as a positive integer. Combining eqs. (31) and (33), the energy spectrum for the Schrodinger equation in $d$-dimensions can be easily obtained:
$$E_n=\frac{-mk^2}{2\hbar^2\left(n+\frac{d-1}{2}\right)^2}.$$
where ;\ \  $n=j+l_{d-1}.$ \\
Note that this is exactly the same as the one we obtained in Section 2., i.e. eq.(10).\\

In conclusion, we see that Schrodinger equation with potential $1/r$ has an accidental degeneracy in any arbitrary dimension. The corresponding spectrum can be found by the representation of its degeneracy group, that is $SO(d+1)$ in $d$ spatial dimensions.\\

\end{document}